\documentclass[12pt,preprint]{aastex}
\newcommand{\pref}{\protect\ref}
\newcommand{\solrad}{\ifmmode{R}_{\rm S}\else${R}_{\rm S}$\fi}
\newcommand{\solmas}{\ifmmode{M}_{\rm S}\else${M}_{\rm S}$\fi}

\newcommand{\ctn}{\ifmmode\kappa\else$\kappa$\fi}

\newcommand{\flxu}{$\,$ergs$\,$cm$^{-2}\,$s$^{-1}$}
\newcommand{\velu}{$\,$km$\,$s$^{-1}$}

\newcommand{\term}[2]{\mbox{$\,^{#1}{\rm #2}$}}
\def\term#1 #2/{\mbox{$\,^{#1}{\rm #2}$}}

\def\mathstacksym#1#2#3#4#5{\def#1{\mathrel{\hbox to 0pt{\lower 
    #5\hbox{#3}\hss} \raise #4\hbox{#2}}}}

\mathstacksym\lta{$<$}{$\sim$}{1.5pt}{3.5pt} 
\mathstacksym\gta{$>$}{$\sim$}{1.5pt}{3.5pt} 
\mathstacksym\lrarrow{$\leftarrow$}{$\rightarrow$}{2pt}{1pt} 
\mathstacksym\lessgreat{$>$}{$<$}{3pt}{3pt} 

\renewcommand{\vec}[1]{{\bf #1}}
\newcommand{\cross}{\times}

\newcommand{\jcb}{\ifmmode\vec{j}\cross\vec{B}\else$\vec{j}\cross\vec{B}$ \fi}

\newcommand\figone{
\begin{figure}[!ht] 
\epsscale{0.8}
\plotone{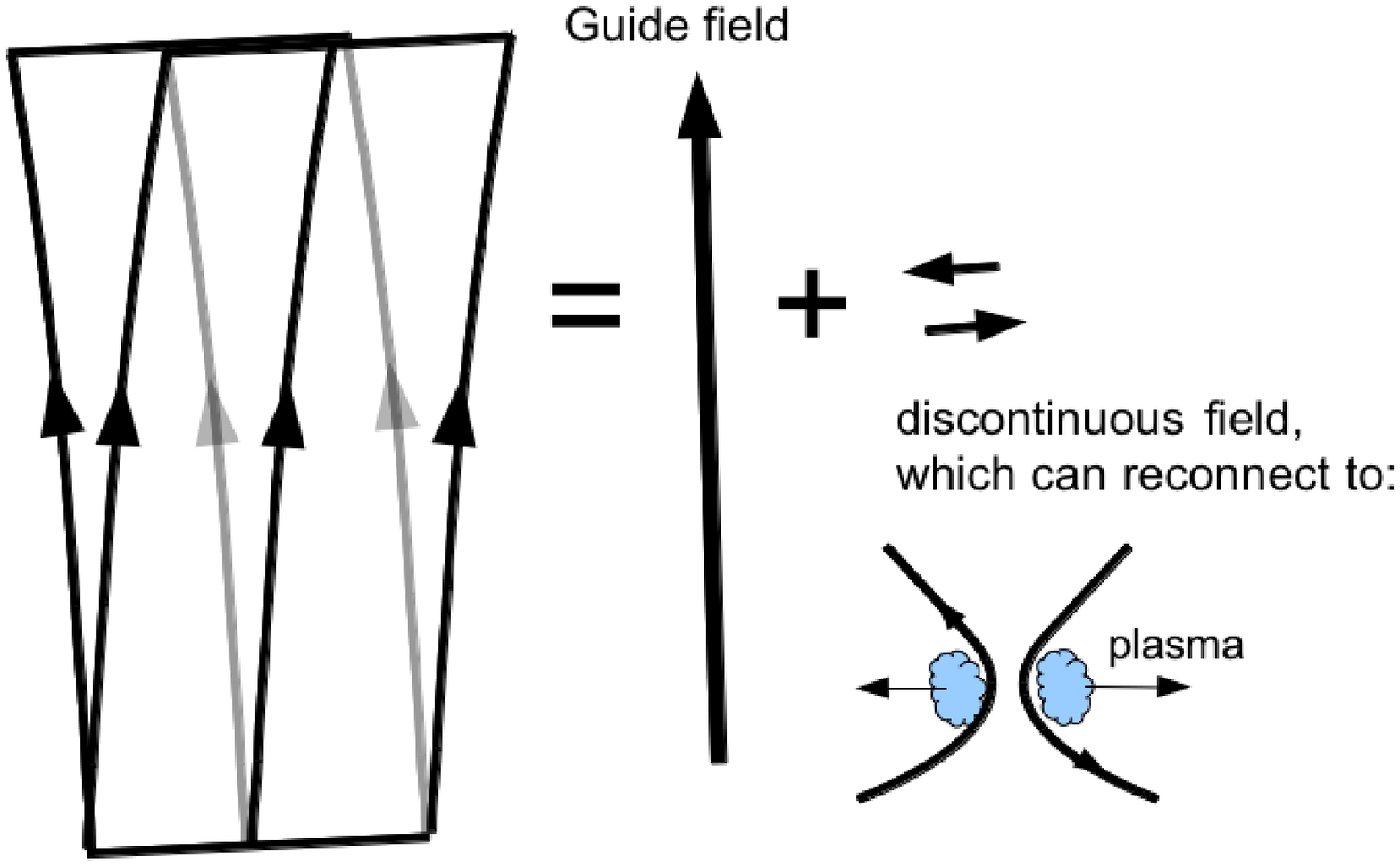}  
\caption{\label{fig:sketch}   A sketch of the envisaged magnetic 
configuration across a current sheet,  plasma is assumed to
be sandwiched between the two magnetic surfaces shown.  
The right hand side of the figure shows a decomposition of the field
across the sheet into the continuous guide field and the discontinuous 
tangential components. The latter can recombine in the manner shown
leading to acceleration of plasma perpendicular to the guide field in
the plane of the sheet.
}
\end{figure}
}

\newcommand\figtwo{
\begin{figure}[!ht] 
\epsscale{0.9}
\plotone{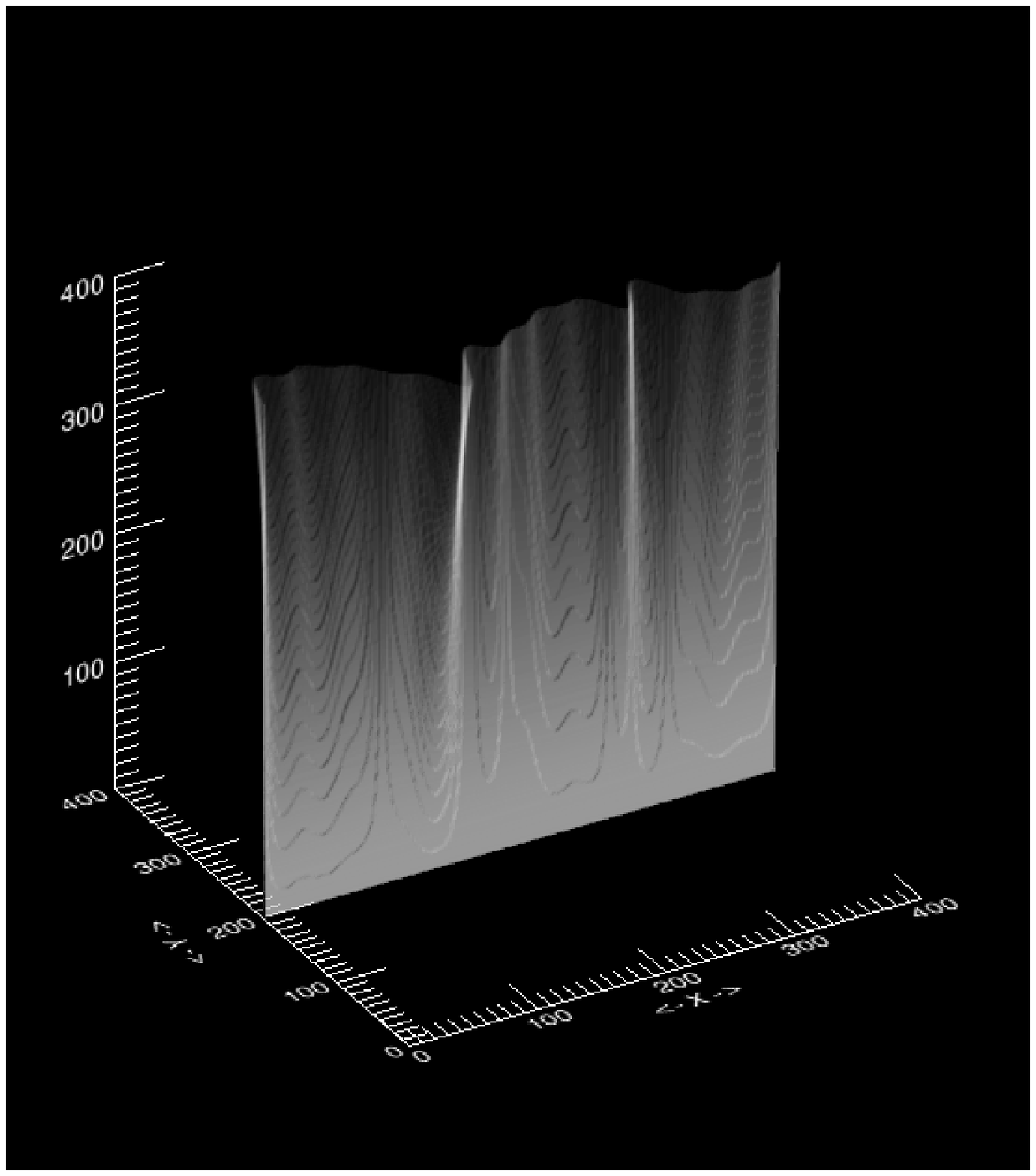}  
\caption{\label{fig:sheet} A warped sheet rendered in 3D.   The
  brightness is proportional to a ``density'' integrated along the
  line of sight, the density is uniform at each ``height'' $z$ within
  the sheet and zero outside it.  The amplitude of the ``warps''
  increases with $z$.  (Axes are in arbitrary units). 
}
\end{figure}
}

\newcommand\figthree{
\begin{figure}[!ht] 
\epsscale{0.9}
\plotone{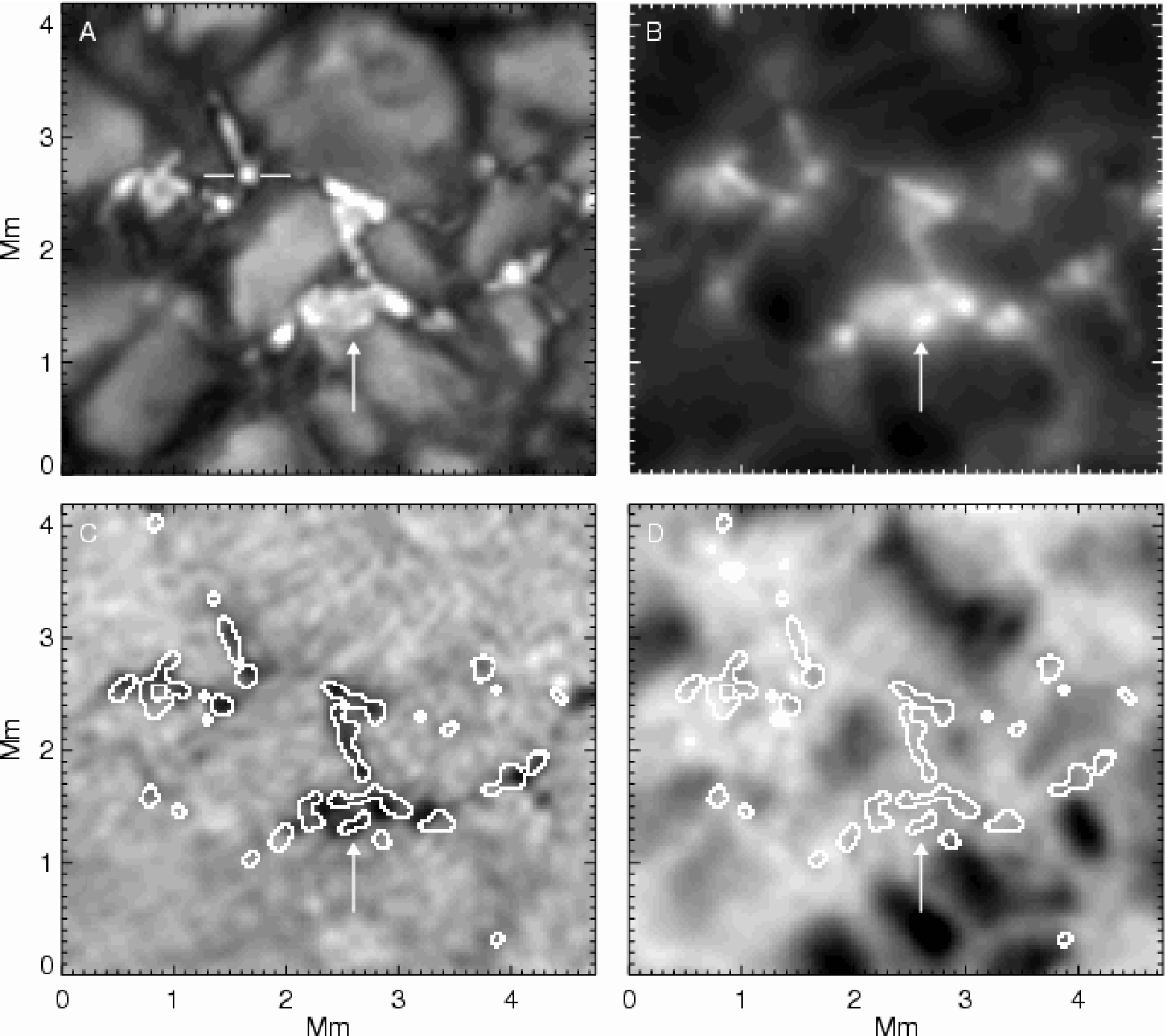}  
\caption{\label{fig:berger} Figure~13 of \citet{Berger+others2004},
  showing  a) G-band filtergram,  b) Ca H-line filtergram,  c) Fe 630.2
  nm magnetogram,  d) Ni I 676.8nm Dopplergram. 
}
\end{figure}
}

\newcommand\figfour{
\begin{figure}[!ht] 
\epsscale{0.9}
\plotone{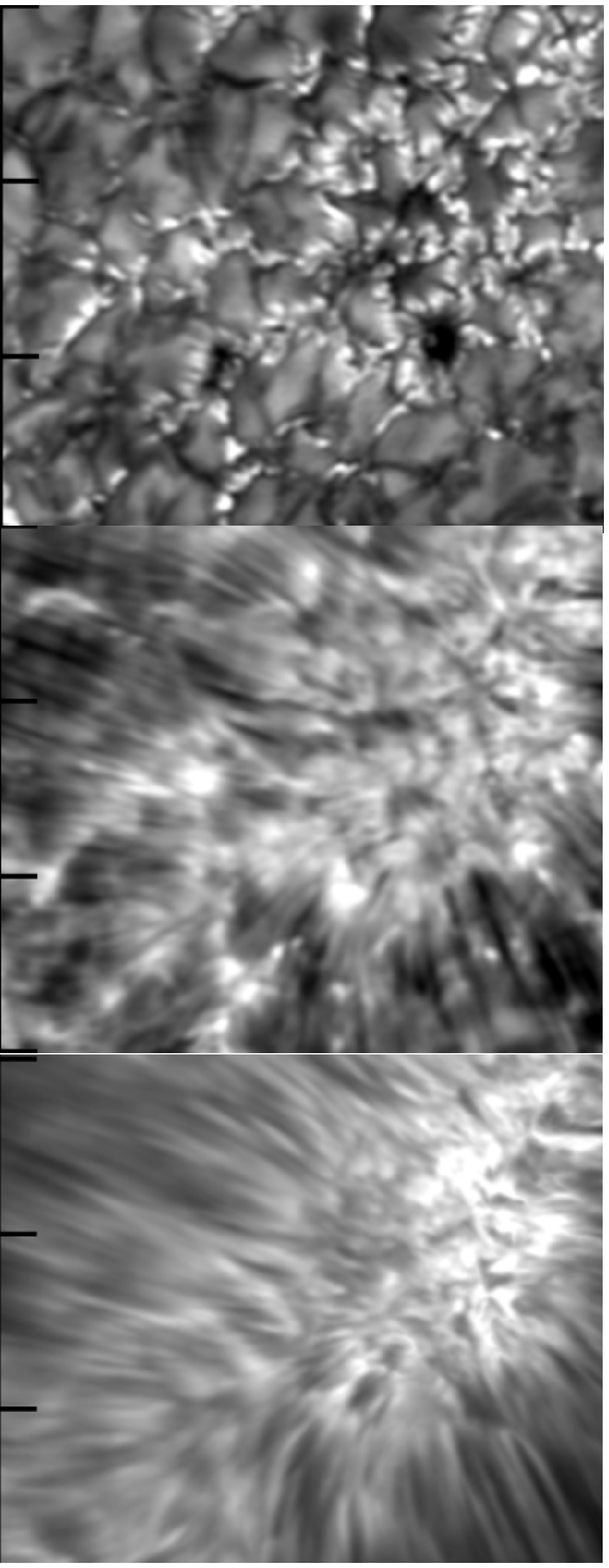}  
\caption{\label{fig:jaime} Images of pores in the 
Ca II 854.2 nm line, obtained with the 
Swedish Solar Telescope, extracted from \citet[][fig
1.5]{delaCruz2010}.  The upper panel shows photospheric 
wing images including granulation and the tortuous sheets of bright
emission associated with the magnetic network.  The middle panel shows the low
chromosphere with reverse granulation, and some fine chromospheric 
fibrils which have some opacity at this wavelength.  The bottom panel
is a line center image.  The FOV is $18\arcsec\times 15\arcsec$
($13\times11$ Mm).  
}
\end{figure}
}

\newcommand\tableprop{
\protect\begin{deluxetable}{llll} \label{tab:spiculeprops}
\renewcommand{\baselinestretch}{1.0}
\tablecaption{Typical spicule properties}
\tablehead{parameter & unit & ``classical''& ``Type II'' }
\startdata
length &Mm & 6.5-9.5 (H$\alpha$) & $\sim5$ (Ca II H) \\
``diameter'' &Mm & 0.4-1.5 & ``very thin'', (down to 0.2 Mm?)\\
lifetime &minutes & 5-20  & 0.5-2  \\
apparent upward velocity &\velu & 20 & 40-200, sometimes $> 250$\\
apparent horizontal motion &\velu{} & usually $<5$ & $\sim 20$ \\
linewidths (FWHM) &\velu{} & 20-40 & n/a \\
number on the Sun & & $\gta10^6$ & $2\times10^7$$^\dag$\\
\\
\enddata 
\tablecomments{Sources: \citet[][classical]{Beckers1972},
  \citet[][type II]{dePontieu+others2007}.  $^\dag$Estimate 
from \citet{Judge+Carlsson2010}, see text for other estimates. 
}
\end{deluxetable}
}

\slugcomment{}

\begin{document}

\newcommand{\philemail}{judge@ucar.edu}
\newcommand{\bcemail}{low@ucar.edu}
\newcommand{\aliemail}{ali@nso.edu}

\shortauthors{Judge et al.}
\shorttitle{}


\slugcomment{}

\title{Thermal fine structure and magnetic fields in the solar
  atmosphere: spicules and fibrils}
\author{Philip G. Judge}
\affil{High Altitude Observatory,
       National Center for Atmospheric Research\altaffilmark{1},
       P.O.~Box 3000, Boulder CO~80307-3000, USA; \philemail}
\author{Alexandra Tritschler}
\affil{National Solar Observatory/Sacramento Peak\altaffilmark{2}, P.O.~Box 62, 
       Sunspot, NM-88349, U.S.A.; \aliemail}

\and
\author{Boon Chye Low}
\affil{High Altitude Observatory,
      National Center for Atmospheric Research\altaffilmark{1},
     P.O.~Box 3000, Boulder CO~80307-3000, USA; \bcemail}

\altaffiltext{1}{The National Center for Atmospheric Research is
sponsored by the National Science Foundation}

\altaffiltext{2}{Operated by the %
       Association of Universities for Research in Astronomy, Inc. (AURA), %
       for the National Science Foundation}

     \begin{abstract} The relationship between observed structures in
       the solar atmosphere and the magnetic fields threading them is
       known only for the solar photosphere, even then imprecisely.
       We suggest that some of the fine structure in the more
       tenuous chromosphere and corona- specifically some populations
       of spicules and fibrils- correspond to warps in 2D sheet-like structures,
       as an alternative to conventional interpretations in terms of 
       tube-like structures.  The sheets are perhaps related to
       magnetic tangential discontinuities, which Parker has argued
       arise naturally in low-$\beta$ conditions. Some consequences of this
       suggestion, if it can be confirmed, are discussed.
       \end{abstract}


\section{Introduction}
\label{sec:introduction}

Unlike data obtained in the laboratory, 
astrophysical data, with a few
important exceptions, are principally obtained by remote sensing.  Nevertheless with 
ingenuity, physical models and luck (the existence of
stars in clusters, standard candles, etc.),  
astrophysics has even been able to challenge fundamental
physics, by revealing the solar neutrino problem, for example.   
Many situations exist however where the physical system observed is
too complex for observations to permit clear-cut conclusions.  In these cases we are armed 
with 
Occam's razor -- the principle that says we should take the simplest
physical picture compatible with the
data -- in order to make progress.  The Sun's
atmosphere, from photosphere to corona,  is one area where Occam's
razor provides a useful
first approach.  But this astrophysical system is highly non-linear, energy transport
is anisotropic and it is strongly 
driven by turbulent convection.  Understanding the  transport of mass, momentum and 
energy within the solar atmosphere and outwards into the heliosphere 
is manifestly important for life on earth. It is also
an astrophysical object uniquely suited to the direct 
observational 
study of naturally occurring magnetized plasmas.

In this {\em letter}
we address the simple question:
are straw-like fine structures seen in the chromosphere and corona 
essentially narrow 
magnetic flux tubes (1D-like) or
something else? Expressed more broadly, 
what is the relationship between the magnetic fields threading the
solar chromosphere and corona, and the observed fine structures?

\section{A case study of low-$\beta$ solar structures: spicules and fibrils}

The solar limb seen in strong (i.e. chromospheric ) spectral lines
reveals dynamic, straw-like structures called
``spicules'', 
famously reported in early visual
spectroscopic observations \citep{Secchi1877}, captured in photographs
and visually observed by \citet{Roberts1945}, quantified by
\citet{Beckers1969,Beckers1972}, and now presented in exquisite detail in
data from the seeing-free platform of the Hinode spacecraft
\citep{dePontieu+others2007}.  Some spicules have been linked with 
the fine structures seen on the solar disk, called chromospheric ``fibrils''  - recent examples of
which are shown in Fig.~\pref{fig:jaime} - associated with 
underlying magnetic flux concentrations 
\citep{Langangen+others2008,Rouppe+others2009}. 
In 1973 Skylab revealed coronal 
plasma organized  into ``loops'' over active regions. The loops were 
envisaged as plasma entrained in
tubes of magnetic flux along which plasma flows under low-$\beta$
conditions ($\beta$= gas pressure/magnetic pressure).
The ``tube'' or ``straw'' picture has become a generally accepted
picture of the low-$\beta$ regions of the Sun's atmosphere -- the magnetized
chromosphere, corona, and even the penumbrae of sunspots.
STEREO observations of active region coronal loops confirm that, on
observable scales ($> 1-2$Mm), the tube picture appears reasonable
\citep{Aschwanden+others2008}.  Yet the plasma inside 
such tubes must consist of 
unresolved ``strands'' whose physical nature is poorly known \citep[e.g.][]{Klimchuk2009}.

\subsection{A hypothesis}

The literature on spicules assumes explicitly or implicitly that these
chromospheric phenomena correspond to plasma within magnetic tubes or straw-like
structures.  In Sterling's (2000) review, we find ``All of the
simulations begin with some form of deposition of energy in the
photospheric or chromospheric portion of a magnetic flux tube which
extends from the photosphere into the corona.''\nocite{Sterling2000}.
We find in Wikipedia: ``...a spicule is a dynamic jet of about 500 km
diameter in the chromosphere of the Sun... their mass flux is about 100
times that of the solar wind.''  While not a formal scientific source,
Wikipedia nevertheless summarizes the current paradigm: we must
understand spicules in terms of field-aligned flows which are more of
a 1-D than 2- or 3-D cool structure embedded in the magnetoplasma.

Noting that no steroscopic observations of  spicules
are yet possible, which will decide the issue once and for all, we hypothesize that at
least some spicules are the {\em observational manifestations of 2D sheet
  structures, physical counterparts of mathematical magnetic
  tangential discontinuities \citep[e.g.][]{Parker1994}}. The spicules
would then be analogous to
the fluted parts of curtains.  Parker argues that as a low-$\beta$ system tries to
maintain its equilibrium and magnetic topology,  tangential
discontinuities will form naturally where 
the magnetic field direction, but not its magnitude,
changes across the sheet.  Following \citep{Parker1988}, we envisage
sheets which have a dominant ``guide'' magnetic field component, the
angles between the discontinuous magnetic field vectors being small (see
Figure~\pref{fig:sketch}).

Plasma may naturally accumulate in such sheets, either by the heating
associated with the dissipative relaxation of the sheet (increasing
the pressure scale height and filling the sheet with plasma from
below) and/or as different magnetic flux bundles are brought together, via the
convective flows \citep{vanBallegooijen+others1998}.
Fig.~\pref{fig:sheet} shows how a smoothly, harmonically fluted or
warped sheet, containing optically thin emitting plasma, can produce
what appears to be a straw-like structure, just from a line of sight
integration.  The equation for this particular sheet in Cartesian
coordinates is simply
$$
y= f(x,z) = 0.01\  (e^{z/z_s}-1)\ \sum_{j=0}^5 \frac{1} {j+1} \sin {(2\pi j  \frac{x}{x_s})} ,
$$ where $x_s$ and $z_s$ are arbitrary characteristic scales, we chose
values of 0.6876 and 0.5, respectively.   No
``tube'' or ``straw'' exists in this configuration. 

The motivations behind the sheet hypothesis are twofold.
First, the highest angular resolution images of the underlying
photospheric magnetic structures (line-of-sight flux density and/or
proxies such as the G-band) show few obvious ``flux tubes'', instead
they appear as fluted sheets
\citep{Berger+others2004,Riethmueller+others2010}.  Figure 10 of
Berger's paper, reproduced in Figure~\pref{fig:berger}, shows a region
of decaying active network\footnote{``Classical'' and ``type II''
  spicules are more prevalent away from active network.  The figure
  was selected here in spite of this because it is one of the clearest
  examples of data revealing the nature of magnetoconvection.}.  In
locally unipolar magnetic regions, such sheet-like structure
must continue into the overlying chromosphere.   In the above equation
the sheet displacements $y$ grow with height $z$ to mimic the combined
growth
of Alfv\'enic fluctuations and natural development of sheet structures. 

This leads to the second motivation, 
magnetic tangential discontinuities must develop in the
low-$\beta$ environment
above the photosphere according to Parker's magnetostatic theorem
\citep{Parker1994,Low2010,Janse+Low+Parker2010}.  Coronal magnetic
fields are braided into complex 3D topologies as their footpoints are shuffled untidily in the convective turbulence of the high
$\beta$ photosphere. This theorem makes the basic point that preserving the
coronal field topologies, under conditions of high electrical conductivity,
naturally drives the fields to equilibrium states embedding such
surfaces of discontinuity.

Our proposal is simply 
that magnetic structures, 
arising naturally from the fluted sheet 
configurations commonly measured in the
photosphere (Fig. \pref{fig:berger}), and subject to the formation of
tangential discontinuities, extend into the 
chromosphere and corona, where they may explain some of the fine
strucure
seen there.

\subsection{Spicules and fibrils as straws}

While almost universally accepted, explicitly or implicitly, 
the straw picture has some
difficulties. 
 How can large numbers of the observed long, thin 
chromospheric fibrils (Fig.~\pref{fig:jaime}) and spicules
\citep{dePontieu+others2007} be formed as tube-like straws from
the kind of magnetic sources found in the photosphere
(Fig.~\pref{fig:berger})?   In this picture, observed kinematics must
be interpreted as real fluid flow- how then are Mach 10-20 flows
suggested by properties of 
spicules of type II generated in the chromosphere (see below)? 

\subsection{Spicules and fibrils as sheets}

We recall known properties of spicules.  Spicules are defined
empirically, but 
owing to problems of atmospheric seeing, the relation
between ``classical'' spicules observed from the ground and the far more
dynamic and finer scale Hinode ``types I and II'' spicules is as yet unclear
\citep{Pasachoff+others2009}.  Nevertheless we summarize observed properties in
Table~\pref{tab:spiculeprops}.  Here we focus 
focus on the ``classical''
spicules of Beckers and the ``type II'' spicules observed with the SOT
instrument on the Hinode spacecraft, of \citet{dePontieu+others2007}.
We do not discuss ``type I'' spicules described by de Pontieu et
al.\footnote{Type I spicules have a satisfactory explanation in terms
  of 1D field aligned recurrent up/down flows driven by work done on
  network flux concentrations by granular motions
  \citep{Hansteen+others2006}. Their material falls back down and so 
  does not ``reach'' the
  corona, and they are apparently rare in coronal holes, where the classical
  spicules and type IIs are most obvious.  Type I spicules cannot
  all correspond  to the ``classical'' spicules, which are most obvious
  in coronal holes, even though some authors equate classical and type
  I spicules 
\citep{Martinez-Sykora+Hansteen+Moreno-Insertis2010}.}. 

What do the best constrained spicule data -- those for type II spicules -- imply
if all such spicules are interpreted as warped sheets?  At any time
there are $\approx6\times10^6$ granules on the Sun
\citep{Namba+Diemel1969}, and, as listed in the table and discussed
further below, $\sim2\times10^7$ type II spicules
\citep{Judge+Carlsson2010}.  With these data we would need on average
4 warps per granule, or 1600 per 20Mm diameter supergranule. Since
spicules are gathered along network lanes, this requires $\lta 20$
warps per Mm along a supergranular lane, or a length scale for the
warps of $\gta 50$ km.  (The inequalities arise because more than one
warped sheet can exist per unit length of the lane.  The same
statistics apply of course to the ``straw'' picture''.)

Lifetimes of type II spicules are on the order of 35-45s, and lie within 
10 - 60s in a Gaussian-like distribution \citep{dePontieu+others2007}.  Granular lifetimes,
defined as the time taken for an entire granule to lose its identify,
are $\approx10$ minutes.  Horizontal granular flows are $v \sim 1-2$
\velu{}, and near downflow lanes, characteristic dynamical
length scales are
$\ell \sim 120$ km.  It seems likely that we should expect to see
dynamical time scales $\ell/v\sim1-2$ minutes at the base of warped
sheets.

Unlike straws, warps do not {\em require} apparent motions to be
attributed to fluid flow, since they result from line of sight
integrations.  As the photospheric field is buffeted by convection,
the fine structures come and go as the plasma along each line of sight
evolves in response to the driving. The apparent motions along type II
spicules are only upward.  In the warp picture this can occur if the
phenomenon is driven mostly from below.  Magnetic stresses propagate
upwards at the Alfv\'en speed, which increases with height.  The
combination of a vertical gradient in Alfv\'en speed, a driver from
below, and optically thin emission, will tend to produce upward
apparent motion along warps that is not related to flows.  If warps
are also influenced from above, and/or more than one warp evolves
along a given line of sight, then downward propagation would be
seen, and interference may produce extremely fast apparent speeds.  In
this regard de Pontieu and colleagues note that type II spicules
``often disappear over their whole length within one or a few time
steps (5-20s)'', and some have apparent velocities $>250$ \velu.  
\citet{dePontieu+others2007} also note that ``a
significant number [of type II spicules] appear to be slower during
[the] very short initial phase and seem to accelerate as they reach
greater heights.''  This behavior is also naturally expected in the
warp picture, as the warps propagate upwards with the increasing
Alfv\'en speed.

The hydromagnetic effects discussed above are still rudimentary but
they merit further theoretical investigation. As a tangential
discontinuity develops on a flux surface the strength of the current
singularity is greatest at the intersection of that surface with the
lowest, densest parts of the atmosphere \citep{Low1990}.  Current
dissipation there may result in upward motion and magnetic
disturbances everywhere on the surface of this continuity, that then
appear enhanced where the surface is fluted.

The magnetic fields of spicules have been explored using polarimetry
\citep{Lopez-Ariste+Casini2005,Centeno+Trujillo+Ramos2010}, from which it is clear that the field
is largely oriented along the apparent axis of the spicule.  This is no
problem for the warp picture in that tangential discontinuities formed
in the Parker picture (as opposed to the ``between tubes'' picture of
\citealp{vanBallegooijen+others1998}) have a dominant guide field
within the plane of the sheet.

\subsection{Dynamics and  connections to the corona}

EUV and optical imaging spectroscopy has revealed blueshifted features
whch have been ascribed indirectly to type II spicules.  
Weak coronal blueshifted
emission has been found which some authors have identified with 
a hot component of the assumed outflow from type II
spicules, since the velocity distributions of
spicules seen at the limb corresponds approximately to the observed
Doppler shift distribution of coronal lines seen against the solar
disk \citep{dePontieu+others2009}.
In this picture, upflowing spicular plasma is assumed to be heated to
$>10^6$ K and the continued upflow is seen directly in blueshifted
coronal lines. Disk counterparts of type II spicules have been 
sought as absorption features in H$\alpha$. A class of rapid blue-shifted events
\citep[RBEs][]{Langangen+others2008,Rouppe+others2009}
has been proposed as the type II counterpart.  Langangen et al. 
note that ``however, the magnitude of the measured Doppler velocity is
significantly lower than the apparent motions seen at the limb".  The
latter authors too find Doppler shift velocities mostly at or below 50 \velu.
\citet{dePontieu+others2010} examine H$\alpha$ and EUV data from
Hinode, finding correlations of RBEs with short-lived brightenings in
a wide range of EUV lines observed.  

This tidy picture of upflowing, cool spicular plasma, that is heated
to 10$^5$ and $10^6$K, explaining the disappearance of the spicule
itself and the appearance of blue-shifted emission in, e.g., He II and coronal
lines, has a basic physical problem. The proposed heating, required to
make the plasma visible as EUV emission, means that the plasma is
over-pressured. The huge additional pressure will broaden velocity
distributions as the plasma dynamically evolves on sound crossing
timescales of a minute or less and radiates He II and coronal lines.
In this picture, therefore, {\em different,
  broader} velocity distributions in the hotter plasmas should be expected. The energy
required to drive such flows is potentially enormous.
\citet{Judge+Carlsson2010} estimated that type II spicules,
interpreted as flows, carry a few times $10^7$ \flxu{} locally in
enthalpy flux.  This is on the order of the heating rate required to
sustain the entire chromosphere
\citep{Vernazza+Avrett+Loeser1981}. But the associated kinetic energy
flux is an order of magnitude higher since the flows have mach numbers
of 10 or so.  Can the apparent 100 \velu{} motions can really be
field-aligned flows?

The scenario proposed by de Pontieu and colleagues may have other
issues. \citet{Rouppe+others2009} find $10^5$ RBEs on the
Sun at any given time, and claim that this is commensurate with the
number of observed type II spicules (1-3 per linear arcsecond along
the limb).  However, the number of type II
spicules on the Sun is probably far larger (see Table 1).  The
estimate of $10^5$ is below even Beckers' (1972) estimate of $10^6$
``classical'' spicules at
zero height, and would correspond to the number of spicules estimated
by Beckers' eq.~(2) at 4000 km above the limb.  This would be a
surprising result given the superior quality of the Hinode data.  The
much larger number of $2\times10^7$ type II spicules given in the
table, from \citet{Judge+Carlsson2010}, was derived by comparing Monte
Carlo radiative transfer calculations with observations.  
This significant difference probably depends on
where in radius one decides to count limb spicules per unit length
along the limb, and contrast thresholds. We regard the synthesis
approach of Judge and Carlsson as more reliable.  We can only
speculate that perhaps RBEs are not then type II spicules, but are
more energetic events more easily seen on the solar disk.

\section{Discussion}

Our proposal extends an earlier idea of
\citet{vanBallegooijen+others1998}, who
suggested that spicules form in between coherent bundles of
photospheric flux as they expand in height, and they interpret
the apparent velocities as real flows.  In our work, we recognize that
current sheets can also form within otherwise continuous 
configurations \citep{Parker1994} and 
we propose that slow modulation of warped sheets can lead to 
supersonic apparent motion seen in off-limb data.
On balance, it seems reasonable to identify some of the fine structure
seen in the cool low-$\beta$ component of the solar atmosphere with
warped current sheets.  

There are several important consequences if indeed significant 
numbers of spicules correspond to
warped current sheets.  
Firstly, such an intepretation presents the prospect of subjecting 
Parker's theory of tangential discontinuities
\citep{Parker1988,Parker1994} to observational scrutiny.    Secondly, as the apparent 
speeds are a mixture of the phase speeds of warps 
(long lines-of-sight) and Alfv\'en speeds, previous estimates of mass
flux ($\sim
100\times$ that of the wind, \citealp{Beckers1972,
Athay+Holzer1982}), which assume that the apparent speeds are real flows,  would be 
too large, re-opening the debate as to the
contribution of spicules to the supply of mass to the corona and wind.
Thirdly, the 2D geometry is important as the sheets present a far
larger area of interaction between cool and hot material than is
apparent from using the straw interpretation, enhancing processes which
perhaps
can help explain the unknown energy balance and peculiar observed 
properties of the lower transition region
\citep{Athay1990,Judge2008}. 

The association of warped tangential discontinuities with spicules may
naturally explain the broad linewidths always observed above the solar
limb, even (perhaps especially) in ``unipolar'' regions.  The sheets
exist as changes in direction of the field are needed to accommodate
force balance and topological constraints.  \citet{Parker1988}
proposed that, at least in active regions, relaxation of 
the changes in field direction
may account for coronal heating, even when the changes are
small fractions of a radian.  The significance of this is that if the
current sheets are continually jostled then forced reconnection of the
non-vertical component, involving a sudden reduction of the angle
between the magnetic fields across the sheet, may be a ubiquitous
process.  The reconnecting component of the field naturally leads to 
flows in direction perpendicular to the guide field, perhaps explaining the linewidths,
as well as supplying some heat to help keep the sheets replenished with plasma.

Even if it is later proven that warped sheets are inconsistent with
data, our suggestion highlights a basic problem: how can one get
straw-like structures out of the fluted sheet fields that appear to
dominate the photospheric network at the current observational limit?
If straws, why is the observed distribution of widths so small and at
the resolution limit of current instrumentation?  
Warps have an advantage in that one need not perform theoretical
contortions to make thin straws out of the observed photospheric sheets.
The problem then becomes one of finding a natural 
physical explanation, analogous to Parker's current sheet theory, to
explain how such thin, essentially 1D structures must naturally arise.

Still unanswered is the important question: how does the Sun make such
long cool structures with lengths say 40$\times$ the hydrostatic pressure scale height
\citep[e.g.][]{Sterling2000}?  This issue is exacerbated with the
realization that a pressure gradient that is steeper than hydrostatic 
is the only force that can accelerate upward flows along a 1D flux
tube.   
\citet{Martinez-Sykora+Hansteen+Moreno-Insertis2010} present an
example of a 
type II spicule candidate taken from a 3D simulation of horizontal flux emerging
into pre-existing vertical granular fields, in which 
Lorentz forces squeeze the chromosphere horizontally 
leading to  vertical deflections of plasma.  The
upward components resemble properties of type II spicules, of
the ``straw'' variety.  The model appears promising, but may have
potential problems. Extrapolation of 
the rate of occurrence of their numerical spicule 
produces just $10^4$ such events on the Sun, 
far smaller than estimated above.  The upwardly
accelerated cool plasma also
expands with height, perhaps in contradiction with observations. 
While these authors point to the role of a current sheet in their
numerical experiment, it should not be confused with the scenario we 
propose.  We emphasize more generally that the magnetic environment in 
which spicules are created - the constrained global field topology and the tendency 
for current sheets to form and dissipate-  may naturally explain
spicule-like structures in terms of sheets.  
The work of \citet{Martinez-Sykora+Hansteen+Moreno-Insertis2010}
is important in producing 
events that produce straw-like spicules, to be further studied and 
tested observationally. We suggest that new
numerical models that deal directly with sheet-like processes and 
structures, should also be developed. 

\section{Conclusions}

Sometimes our eyes deceive us:
Forests appear solid from a distance; spiral nebulae, once believed to
be gas clouds, were first resolved into stars by \citet{Hubble1929}.
Warped current sheets present an alternative physical hypothesis for
the nature of spicules  that is 
worthy of further study.  The tube vs. warps picture is not an
either/or proposition- both have advantages and disadvantages when 
pitted against critical observational data.  Further studies of both
hypotheses, 
observational and theoretical, should help
our continued understanding of this difficult area of research.

\acknowledgments We thank Christina Cohen for 
insightful comments on the manuscript, Jaime de la Cruz Rodriguez
for his figures, and the referee for bringing the recent work of 
Mart\'inez-Sykora and colleagues to our attention.


\tableprop
\figfour
\figone
\figtwo
\figthree

\end{document}